# Topological spin transport of photons: "magnetic monopole" gauge field in Maxwell equations and polarization splitting of rays in periodically inhomogeneous media

K.Yu. Bliokh[1,2*] and V.D. Freilikher[2,3]

[1]*Institute of Radio Astronomy, 4 Krasnoznamyonnaya st., Kharkov, 61002, Ukraine*
[2]*Department of Physics, Bar-Ilan University, Ramat Gan 52900, Israel*
[3]*Complex Photonic Systems, Department of Science and Technology and MESA+ research institute, University of Twente, P.O. Box 217, 7500 AE Enschede, The Netherlands*

Topological spin transport of electromagnetic waves (photons) in stationary smoothly inhomogeneous isotropic medium is studied. By diagonalizing photon kinetic energy in Maxwell equations we derive the non-Abelian pure gauge potential in the momentum space, which in adiabatic approximation for transverse waves takes the form of two Abelian potentials **U**(1) corresponding to magnetic-monopole-type fields. These fields act on circularly polarized waves resulting in the topological spin transport of photons. We deduce general semiclassical (geometrical optics) ray equations that take into account a Lorentz-type force of the magnetic-monopole-like gauge field. Detailed analysis of rays in 3D medium with 2D periodic inhomogeneity is presented. It is shown that rays located initially in the inhomogeneity plane experience topological deflections or splitting that move them out from this plane. The dependence of the rays' deflection on the parameters of the medium and on the direction of propagation is studied.

## 1. Introduction

The remarkable property of Maxwell equations is that being the main tool for description of classical electromagnetic waves, they also represent the quantum relativistic Schrödinger type equations for photons if the permittivity of a (inhomogeneous) medium and wave polarization are considered as the potential energy and photon spin respectively. That is why investigating electromagnetic waves one can discover some fundamental properties of relativistic quantum particles with spin 1. For example, the geometrical phase introduced by Berry in the general form [1], had been first described theoretically by Rytov and Vladimirskii [2,3] (see also [4]) in terms of the polarization evolution for electromagnetic waves. Experimentally Berry phase was observed first also for electromagnetic waves [5]. In what follows we assume the equivalence of the notions "photon" and "electromagnetic wave" as well as theirs "spin" and "polarization" respectively.

In recent years progressively increasing attention is given to the phenomena of topological spin transport of quantum particles (see, for instance [6–13] and references there in). These effects are closely related to the general notion of Berry phase and thus should also show up in the electromagnetic wave transport. Indeed, one can find the first description of polarization transport phenomena in works by Zel'dovich et al. [14,15], where the deflections of circularly polarized light beams (so-called "optical Magnus effect") had been predicted theoretically and observed experimentally in fibers. Later it was shown ([16–20]) that this effect represents *the topological spin transport of photons*. The essence of the phenomenon is that in an inhomogeneous medium a polarization (spin) current arises that is directed orthogonally to the local wave vector. It can lead to the topological splitting of a ray of mixed (non-circular) polarization into two circularly polarized rays. As it was demonstrated in [18,19], this effect is completely analogous to the anomalous and spin Hall effects in solids (see [6–8,10,11,13]).

---

[*] E-mail: k_bliokh@mail.ru

In this paper we present a rigorous derivation of the gauge field for electromagnetic waves (photons) directly from the Maxwell equations using the standard diagonalization procedure (see [7,12,13]). As the result, a $\mathbf{U}(3)$ non-Abelian pure gauge potential arises in the momentum space, which is reduced in adiabatic approximation to $\mathbf{U}(2) \times \mathbf{U}(1)$ non-trivial potential. Here $\mathbf{U}(2)$ potential corresponds to the transverse electromagnetic waves and takes the form of product of Abelian gauge potentials $\mathbf{U}^2(1) = \mathbf{U}(1) \times \mathbf{U}(1)$ with magnetic monopole type gauge fields of opposite signs. Thus, the initial 3D vector equation falls into two scalar Helmholtz equations with additional gauge $\mathbf{U}(1)$ potentials in the momentum space. These equations describe two eigen states of a photon with helicity $\pm 1$ (in another words, right-hand and left-hand circularly polarized electromagnetic waves), that undergo the topological spin deflections in inhomogeneous dielectric media. Adiabatic approximation used is equivalent (for stationary medium) to the semiclassical, or geometrical optics approximation, i.e. it implies smoothness of the medium. The effect of topological polarization deflection in a 3D medium with periodic two-dimensional inhomogeneity is studied. We show that right-hand and left-hand circularly polarized rays located in the inhomogeneity plane can be deflected in opposite directions orthogonally to this plane due to topological polarization transport. The deflections strongly depend on the angle of wave propagation, and for the square lattice has a sharp peak at the angle $\pi/4$. For samples long enough there are also additional sharp peaks, which correspond to the irrational values of the tangent of the angle. If the ray has mixed (non-circular) polarization, the topological polarization transport can lead to its splitting into two circular rays or to the depolarization.

## 2. "Magnetic monopole" gauge field in momentum space in Maxwell equations

**2.1. Diagonalization of Maxwell equations and appearance of a gauge potential.** Let us consider a monochromatic electromagnetic wave in a stationary isotropic inhomogeneous medium characterized by the refractive index $n(\mathbf{R}) = \sqrt{\varepsilon(\mathbf{R})}$ ($\varepsilon$ is the dielectric constant of the medium that in general case depends on the wave frequency $\omega$, and $\mathbf{R}$ is the coordinate vector). Maxwell equations for the wave electric field $\mathcal{E}$ have the form

$$\left[ \text{curl curl} - k_0^2 n^2 \right] \mathcal{E} = 0, \tag{1}$$

where $k_0 = \omega / c$. If we introduce the dimensionless momentum differential operator

$$\mathbf{p} = -i k_0^{-1} \frac{\partial}{\partial \mathbf{R}}, \tag{2}$$

the equation (1) takes the form

$$-\mathbf{p} \times (\mathbf{p} \times \mathcal{E}) - n^2 \mathcal{E} = 0. \tag{3}$$

(For monochromatic electromagnetic waves and Maxwell equations it is more convenient to define the momentum with factor $k_0^{-1}$ rather than with Planck constant $\hbar$.)

Eq. (3) is a relativistic Schrödinger type equation:

$$\hat{H} \mathcal{E} = 0, \tag{4}$$

with matrix-valued Hamiltonian differential operator (here and further on we mark matrices by hats)

$$\hat{H}(\mathbf{p}, \mathbf{R}) = \hat{I} \left[ p^2 - n^2(\mathbf{R}) \right] - \hat{Q}. \tag{5}$$

Here $Q_{ij} = p_i p_j$, $\hat{I}$ is $3 \times 3$ unit matrix, $p = |\mathbf{p}|$, and $-n^2$ and $\hat{I}p^2 - \hat{Q}$ play the role of the potential energy and non-diagonal kinetic energy of the electromagnetic wave (photons) respectively. Note that photon Hamiltonian operator in quantum electrodynamics is quite similar to Eq. (5) and has the same non-diagonal part [21].

Obviously, non-diagonal Hamiltonian (5) mixes up different independent eigen states of the wave. To find these eigen states and to describe their evolution we diagonalize the Hermitian Hamiltonian (5) using the general procedure with an unitary matrix $\hat{U}(\mathbf{p})$ (see, for instance, [7,12,13]). Matrix $\hat{U}$ is built of the orthogonal set of unit eigen vectors of matrix $\hat{H}$ and can be chosen as

$$\hat{U} = \begin{pmatrix} \dfrac{p_y}{\sqrt{p_x^2 + p_y^2}} & -\dfrac{p_x p_z}{p\sqrt{p_x^2 + p_y^2}} & \dfrac{p_x}{p} \\ -\dfrac{p_x}{\sqrt{p_x^2 + p_y^2}} & -\dfrac{p_y p_z}{p\sqrt{p_x^2 + p_y^2}} & \dfrac{p_y}{p} \\ 0 & \dfrac{\sqrt{p_x^2 + p_y^2}}{p} & \dfrac{p_z}{p} \end{pmatrix}. \quad (6)$$

With similarity transformation with Eq. (6) the non-diagonal part $\hat{Q}$ of Hamiltonian (5) becomes diagonal:

$$\hat{U}^{-1}\hat{Q}\hat{U} = \hat{\Lambda} \equiv \mathrm{diag}(\lambda_1, \lambda_2, \lambda_3) = \begin{pmatrix} 0 & 0 & 0 \\ 0 & 0 & 0 \\ 0 & 0 & p^2 \end{pmatrix}. \quad (7)$$

One can see that two eigen values of $\hat{Q}$ coincide with each other and are equal to zero, $\lambda_{1,2} = 0$, while the third one is equal to $\lambda_3 = p^2$. It follows from Eqs. (5) and (7) that eigen values $\lambda_{1,2}$ correspond to transverse electromagnetic waves with local dispersion law $p^2 = k_0^2 n^2$ (or $\omega^2 = k^2 c^2$, $k$ is the local wave number), while the eigen value $\lambda_3$ corresponds to the longitudinal wave, which can exist only at resonances where $n = 0$ [22]. In what follows we assume that such resonant points (if they exist) are well off the region of propagation, and longitudinal waves are not excited. The transverse wave state is double degenerated that presents *polarization degeneration* of transverse electromagnetic waves: a phenomenon well-known in geometrical optics of isotropic media [22], and corresponds to the *spin degeneration* of energy levels of relativistic quantum particles. Note also that unitary transformation with $\hat{U}(\mathbf{p})$ is equivalent to the local rotation of the coordinate frame in such a way that $z$-axis in each point is directed along $\mathbf{p}$ (it can be seen from the comparison of Eq. (7) with the initial matrix $Q_{ij} = p_i p_j$). Therefore in the new coordinate frame the electric filed $\mathbf{E}$ of the longitudinal state is directed practically along $z$-axis, while the electric field of transverse states lies almost in $(x,y)$-plane (see below).

Transformation (6) being applied to Eq. (4) with Hamiltonian (5) after substitution $\mathcal{E} = \hat{U}\mathbf{E}$ yields:

$$\left[\hat{I}p^2 - \hat{\Lambda} - \hat{U}^{-1}n^2\hat{U}\right]\mathbf{E} = 0. \quad (8)$$

If momentum $\mathbf{p}$ and coordinates $\mathbf{r}$ are ordinary commuting numbers, the potential part of the Hamiltonian (8) is a scalar, $n^2$. However, when electromagnetic wave (or quantum particle) is concerned, the momentum and coordinates are non-commuting quantities that can be represented by differential operators. Therefore in $\mathbf{p}$-representation in Eq. (8) we have

$$\hat{U}^{-1}(\mathbf{p})n^2(\mathbf{R})\hat{U}(\mathbf{p}) = \hat{U}^{-1}(\mathbf{p})n^2\left(ik_0^{-1}\frac{\partial}{\partial \mathbf{p}}\right)\hat{U}(\mathbf{p}) = n^2\left(ik_0^{-1}\hat{I}\frac{\partial}{\partial \mathbf{p}} + k_0^{-1}\hat{\mathbf{A}}\right) = n^2\left(\hat{I}\mathbf{R} + k_0^{-1}\hat{\mathbf{A}}\right), \quad (9)$$

where

$$\hat{\mathbf{A}}(\mathbf{p}) = i\hat{U}^{-1}\frac{\partial \hat{U}}{\partial \mathbf{p}} \quad (10)$$

is a pure gauge non-Abelian potential induced in the momentum $\mathbf{p}$-space by the local gauge transformation $\hat{U} \in \mathbf{U}(3)$ [7,12,13]. In accordance to the gauge field theory and Hamiltonian mechanics, in Eq. (9) $\mathbf{R}$ are *canonical* (or *generalized* in classical mechanics) coordinates that correspond to the "usual" derivatives $ik_0^{-1}\partial/\partial \mathbf{p}$. At the same time, *observable* (i.e. related to 'the center of the particle') coordinates $\hat{\mathbf{r}}$ correspond now to the *covariant* derivatives:

$$\hat{\mathbf{r}} = ik_0^{-1}\frac{D}{D\mathbf{p}} = ik_0^{-1}\hat{I}\frac{\partial}{\partial \mathbf{p}} + k_0^{-1}\hat{\mathbf{A}}. \quad (11)$$

Matrix-valued coordinates $\hat{r}_i$ commute with each other, $[\hat{r}_i, \hat{r}_j] = 0$, because the potential (10) is a pure gauge one, and corresponding field tensor is zero (see [7,12,13], compare also with [20]).

Thus, equation (8) after substitutions Eqs. (9), (10) takes the form

$$\left[\hat{I}p^2 - \hat{\Lambda} - n^2(\hat{I}\mathbf{R} + k_0^{-1}\hat{\mathbf{A}})\right]\mathbf{E} = 0. \quad (12)$$

Although this equation has Hamiltonian of a diagonal form, the potential energy $n^2$ contains a non-diagonal gauge vector potential $\hat{\mathbf{A}}$. This potential describes all non-trivial (i.e. those that are not connected to the current eigen values of the initial Hamiltonian) evolutions of the different polarization states of the wave. Direct calculation of Eq. (10) with Eq. (6) leads to the following expressions for the Hermitian antisymmetric matrices $\hat{A}_{p_i}$:

$$\hat{A}_{p_x} = i\begin{pmatrix} 0 & -\frac{p_y p_z}{p(p_x^2 + p_y^2)} & \frac{p_y}{p\sqrt{p_x^2 + p_y^2}} \\ \frac{p_y p_z}{p(p_x^2 + p_y^2)} & 0 & -\frac{p_x p_z}{p^2\sqrt{p_x^2 + p_y^2}} \\ -\frac{p_y}{p\sqrt{p_x^2 + p_y^2}} & \frac{p_x p_z}{p^2\sqrt{p_x^2 + p_y^2}} & 0 \end{pmatrix}, \quad (13a)$$

$$\hat{A}_{p_y} = i\begin{pmatrix} 0 & \frac{p_x p_z}{p(p_x^2 + p_y^2)} & -\frac{p_x}{p\sqrt{p_x^2 + p_y^2}} \\ -\frac{p_x p_z}{p(p_x^2 + p_y^2)} & 0 & -\frac{p_y p_z}{p^2\sqrt{p_x^2 + p_y^2}} \\ \frac{p_x}{p\sqrt{p_x^2 + p_y^2}} & \frac{p_y p_z}{p^2\sqrt{p_x^2 + p_y^2}} & 0 \end{pmatrix}, \quad (13b)$$

$$\hat{A}_{p_z} = i\begin{pmatrix} 0 & 0 & 0 \\ 0 & 0 & \frac{\sqrt{p_x^2 + p_y^2}}{p^2} \\ 0 & -\frac{\sqrt{p_x^2 + p_y^2}}{p^2} & 0 \end{pmatrix}. \quad (13c)$$

In these expressions $2\times 2$ sectors that consist of 11, 12, 21, and 22 elements describe the evolution of the double degenerated transverse states; the zero component 33 acts on the longitudinal state, while the cross-components 13, 23 and 31, 32 describe the transitions between transverse and longitudinal wave states.

**2.2. Adiabatic reduction and separation of eigen states.** As it was mentioned above, if the resonance points where $n(\mathbf{r},\omega) = 0$ are far away from the region of propagation, longitudinal waves are not excited. Then in smoothly inhomogeneous medium the electric field is almost transverse, $\mathbf{E} \cong (E_x, E_y, 0)$, and the longitudinal component of the electric field, $E_z$ is proportional to the small geometrical optics (semiclassical) parameter [22]

$$\mu = \frac{1}{nk_0 L} << 1 , \qquad (14)$$

where $L$ is the characteristic scale of medium inhomogeneity. Of the same order, $\mu$, are the cross-state terms, which appear in the motion equations from the matrix operator (13). It is known that in this case their contribution to the states evolution is proportional to $\mu^2$ (see, for example in [23]), and, therefore, in the first approximation in $\mu$ one should neglect 13, 23 and 31, 32 components of $\hat{\mathbf{A}}$. Then the initial gauge potential $\mathbf{U}(3)$ is reduced to $\mathbf{U}(2) \times \mathbf{U}(1)$ that corresponds to the *adiabatic* approximation (see, for instance, in [7,12,13,20]). (As it is shown in [13] the adiabaticity condition for a non-localized (propagating) particle with stationary Hamiltonian is equivalent to a semiclassical condition similar to Eq. (14).) Thus the transverse and longitudinal wave states become independent, and equation (12) breaks down into $2\times 2$ matrix equation for the transverse states and a scalar equation for the longitudinal one. Matrix equation for 2-dimensional transverse electric vector field $\widetilde{\mathbf{E}} = (E_x, E_y)$ has the form:

$$\left[\hat{I}p^2 - n^2\left(\hat{I}\mathbf{R} + k_0^{-1}\hat{\mathbf{A}}^{(tr)}\right)\right]\widetilde{\mathbf{E}} = 0 . \qquad (15)$$

Here $\hat{I}$ is $2\times 2$ unit matrix; $\hat{\mathbf{A}}^{(tr)}$ denotes the "transverse" $2\times 2$ sector of the corresponding $3\times 3$ matrices, Eq. (13). In the adiabatic approximation $\hat{\Lambda}^{(tr)} = 0$ and it disappears at conversion of Eq. (12) into Eq. (15).

As it is well-known from the Berry phase and spin gauge field theories, the potential $\hat{\mathbf{A}}^{(tr)}$, that corresponds to double degenerated state, is a $\mathbf{U}(2)$ non-Abelian gauge potential, which can be decomposed into full set of non-commuting Pauli matrices (see, for instance, [24,25,20]). However, as one can see from the explicit form of matrices (13), when it comes to photons, matrix $\hat{\mathbf{A}}^{(tr)}$ is proportional to the single Pauli matrix $\hat{\sigma}_2$:

$$\hat{\mathbf{A}}^{(tr)} = \left(\frac{p_y p_z}{p(p_x^2 + p_y^2)}, -\frac{p_x p_z}{p(p_x^2 + p_y^2)}, 0\right)\hat{\sigma}_2 , \quad \hat{\sigma}_2 = \begin{pmatrix} 0 & -i \\ i & 0 \end{pmatrix} . \qquad (16)$$

Components of $\hat{\mathbf{A}}^{(tr)}$ commute with each other, which means that $\hat{\mathbf{A}}^{(tr)}$ represents Abelian potential $\mathbf{U}^2(1) = \mathbf{U}(1) \times \mathbf{U}(1)$. Obviously, we can rotate the reference frame in such a way that $\hat{\sigma}_2$ transforms into $\hat{\sigma}_3 = \begin{pmatrix} 1 & 0 \\ 0 & -1 \end{pmatrix}$, and potential $\hat{\mathbf{A}}^{tr}$ becomes diagonal. Such uniform rotation is executed by the global substitution $\widetilde{\mathbf{E}} = \hat{V}^{(tr)}\Psi$ with

$$\hat{V}^{(tr)} = \frac{1}{\sqrt{2}}\begin{pmatrix} 1 & 1 \\ i & -i \end{pmatrix} , \qquad (17)$$

which diagonalize equation (15), and therefore breaks it down into two independent scalar equations:

$$H^{\pm}\Psi^{\pm} = 0 , \quad H^{\pm} = \frac{1}{2}\left[p^2 - n^2\left(\mathbf{R} + k_0^{-1}\mathbf{A}^{\pm}\right)\right] = \frac{1}{2}\left[p^2 - n^2\left(\mathbf{r}^{\pm}\right)\right] , \qquad (18)$$

where $\mathbf{r}^{\pm} = ik_0^{-1}\partial/\partial\mathbf{p} + k_0^{-1}\mathbf{A}^{\pm}$ are the operators of the coordinates of 'the photon's center' (covariant coordinates) analogously to Eq. (11). Here we denote $\Psi \equiv (\Psi^+, \Psi^-)$, 1/2 factor is introduced for convenience, and **U**(1) gauge potentials are equal to

$$\mathbf{A}^{\pm} = \pm\left(\frac{p_y p_z}{p(p_x^2 + p_y^2)}, -\frac{p_x p_z}{p(p_x^2 + p_y^2)}, 0\right). \tag{19}$$

Thus, we have reduced the initial *vector* problem of the evolution of a polarized wave to two independent *scalar* equations (18) with additional gauge potentials (19) (compare with scalar and vector geometric optics in [22]).

It worth noticing that two functions, $\Psi^{\pm} = E_x \mp iE_y$ (see Eq. (17)) represent basis of circularly polarized waves. It means that *only circularly polarized states are the independent eigen states of transverse electromagnetic wave* [16–18]. It is quite natural because these solutions are the photon spin eigen states with helicities $\pm 1$ (see, for instance, [21]). Of course we could apply the transformation (17) directly to the 3-dimensional equation (12). In this case transformation matrix $\hat{V}$ is equal to

$$\hat{V} = \frac{1}{\sqrt{2}}\begin{pmatrix} 1 & 1 & 0 \\ i & -i & 0 \\ 0 & 0 & \sqrt{2} \end{pmatrix}. \tag{20}$$

One can see that columns of this matrix are the eigen vectors of the $z$-component of the photon spin operator $\hat{s}_3$ [21]. This fact supports our conclusion that functions $\Psi^{\pm} = E_x \mp iE_y$ correspond to the independent photon spin states.

**2.3. Monopole type field in the momentum space and non-commuting coordinates.**
After **U**(1) Abelian gauge potentials have been found (Eq. (19)), the corresponding field tensors (Berry curvatures) can be calculated directly, which yields:

$$F_{ij}^{\pm} = \frac{\partial A_j^{\pm}}{\partial p_i} - \frac{\partial A_i^{\pm}}{\partial p_j} = \mp e_{ijk}\frac{p_k}{p^3}, \tag{21}$$

where $e_{ijk}$ is unit antisymmetric tensor. Eq. (21) presents the magnetic-monopole-like field, though not in the real but in the momentum space (see, for example, [1,4,7,8,12,13,18–20,26]). Since **p**-space is a 3-dimensional one the filed (21) can be represented as a pseudo-vector:

$$\mathbf{F}^{\pm} = \mp\frac{\mathbf{p}}{p^3}. \tag{22}$$

Easy to see, these fields act (in **p**-space) on the different spin (polarization) eigen states of the photon ($\Psi^{\pm}$) in opposite directions. As is known, such fields cause *spin current* of particles orthogonal to their motion [6–8,10–13,18–20]. When photons are concerned, they lead to the topological *splitting* of right and left circularly polarized beams (photons) [16–20].

In a recent paper [20] non-Abelian spin gauge potential and field have been derived for a relativistic particle with arbitrary spin. For massless particles this field has the magnetic-monopole-type configuration, which is characteristic for Abelian **U**(1) field. As we have shown above for photons, non-Abelian gauge field **U**(2) is indeed decomposed into two independent Abelian **U**(1) fields.

From Eq. (11) with potentials (19) and fields (21), (22) it follows that coordinates of circularly polarized waves do not commute (see [7,13,20]):

$$\left[r_i^{\pm}, r_j^{\pm}\right] = ik_0^{-2}F_{ij}^{\pm} = \mp ik_0^{-2}e_{ijk}\frac{p_k}{p^3}. \tag{23}$$

Important to emphasize that non-zero field tensors (21) and commutation relations appear owing to the adiabatic reduction $\mathbf{U}(3) \to \mathbf{U}(2) \times \mathbf{U}(1)$ in the gauge potential (10), (13). In the general $\mathbf{U}(3)$ case potential (10), (13) is a pure gauge one, and the field tensor, as well as the coordinate commutators, are zero [7,13].

**2.4. Semiclassical ray equations, topological spin transport of photons, and Berry phase.** If the variations of the refractive index are smooth enough, so that the semiclassical condition (14) is satisfied, Hamiltonian equations of motion for photons (ray equations) can be derived in geometrical optics approximation [22], and have the form

$$\dot{\mathbf{p}} = -\frac{\partial H^{\pm}}{\partial \mathbf{R}}, \quad \dot{\mathbf{R}} = \frac{\partial H^{\pm}}{\partial \mathbf{p}}, \tag{24}$$

or

$$\dot{\mathbf{p}} = -\frac{\partial H^{\pm}}{\partial \mathbf{r}^{\pm}}, \quad \dot{\mathbf{r}}^{\pm} = \frac{\partial H^{\pm}}{\partial \mathbf{p}} + k_0^{-1}\left(\mathbf{F}^{\pm} \times \dot{\mathbf{p}}\right), \tag{24a}$$

where dot denotes the derivative with respect to the ray parameter $s$, which is connected with the ray length, $l$, as $dl = nds$. From now on momentum $\mathbf{p}$ is a dimensionless wave vector (not a differential operator as above), $\mathbf{p} = k_0^{-1}\mathbf{k}$, while $\mathbf{r}^{\pm}$ are the coordinates on the rays of circular polarizations. At that, the dispersion relation $p = n$ is valid. Upon substituting Hamiltonian (18) into Eqs. (24) we obtain [16–20]:

$$\dot{\mathbf{p}} = \frac{1}{2}\frac{\partial n^2}{\partial \mathbf{r}^{\pm}}, \quad \dot{\mathbf{r}}^{\pm} = \mathbf{p} \mp k_0^{-1}\left(\frac{\mathbf{p}}{p^3} \times \dot{\mathbf{p}}\right). \tag{25}$$

Eqs. (25) are transformed into standard geometrical optics equations [22] if the term proportional to $k_0^{-1}$ is neglected. This term represents a Lorentz-type force of the magnetic monopole-like field (22) in momentum space, which causes additional deflections of circularly polarized rays: $\mathbf{r}^{\pm} = \mathbf{r} + \delta\mathbf{r}^{\pm}$, where $\mathbf{r}$ is the coordinates on the zero-approximation ray. These deflections can be calculated by the integration of the last term in second Eq. (25) and is presented as the contour integral [16–18]:

$$\delta\mathbf{r}^{\pm} = \mp k_0^{-1}\int_C \frac{\mathbf{p} \times d\mathbf{p}}{p^3}, \tag{26}$$

where contour $C$ is the wave trajectory in the momentum space.

Lorentz-type force similar to that in Eq. (25) also arises at the non-relativistic electron motion with spin-orbit interaction, where it leads to the anomalous and intrinsic spin Hall effects [6–8,10–13]. In our case the force splits the trajectories of right and left circularly polarized waves and, hence, induces photon spin current orthogonal to the vector $\mathbf{p}$. Thus the term that proportional to $k_0^{-1}$ in equation (25) and Eq (26) presents the *topological spin transport of photons*, which can be associated with *intrinsic spin Hall effect for photons* (see [18,19]). One of the manifestations of the Lorentz-type force in Eq. (25), so-called optical Magnus effect, had been discovered in [14,15]. Recently, equations (25) have been derived and analysed in details in terms of geometrical optics and Berry phase formalism in [16,17], and by postulating monopole field (21), (22) in geometrical optics equations in [18,19]. In the present paper we present the first rigorous derivation of the monopole gauge field (21), (22) and transport equations (25) directly from Maxwell equations.

Note, Berry phase can also be deduced from the approach suggested above (see [13,18]). Indeed, the semiclassical phase is determined in the canonical (generalized) coordinates, and for circularly polarized waves equals

$$\varphi^{\pm} = k_0\int_0^{\mathbf{R}}\mathbf{p}d\mathbf{R} = k_0\int_0^{\mathbf{r}^{\pm}}\mathbf{p}d\mathbf{r}^{\pm} - \mathbf{p}\mathbf{A}^{\pm}\Big|_{\mathbf{p}_0}^{\mathbf{p}} + \int_C \mathbf{A}^{\pm}d\mathbf{p} = k_0\int_0^{\mathbf{r}}\mathbf{p}d\mathbf{r} + \int_C \mathbf{A}^{\pm}d\mathbf{p}. \tag{27}$$

Here the connection between observable and generalized coordinates, Eq. (18), as well as the relations $\delta \dot{\mathbf{r}}^{\pm} \perp \mathbf{p}$, Eqs. (25), (26), and $\mathbf{A}^{\pm} \perp \mathbf{p}$, Eq. (19), have been used; $\mathbf{p}_0$ is the initial momentum, term $-\omega t$ in the phase is omitted. First term in Eq. (27) is the ordinary dynamic phase. The second term is the Berry phase, which for closed contours (loops) can be presented as the flux of the field (21), (22) through the surface $S$, which is stretched over the loop $C$ in $\mathbf{p}$-space:

$$\theta_B = \oint_C \mathbf{A}^{\pm} d\mathbf{p} = \int_S \mathbf{F}^{\pm} d\mathbf{s} \; . \tag{28}$$

For non-circularly polarized electromagnetic waves Berry phase leads to the rotation of the polarization plane described theoretically in [2,3] (see also [4]) and observed experimentally in [5].

## 3. Topological polarization splitting of rays in a periodically inhomogeneous medium

As an example of the effects described above we consider the deflection of rays in a two-dimensionally periodic medium. An advantage of this system is that being relatively simple for analytical analysis, it enables one to demonstrate the most general features of the topological polarization (spin) splitting in inhomogeneous media. On the other hand, 2D periodic systems are interesting by themselves because they find ever increasing applications in optronics and nanotechnology (photonic crystals, nanostructured dielectrics, etc.). Note that in one-dimensionally inhomogeneous media, where all rays are flat curves, the topological effects like Berry phase and polarization splitting of rays (26) never occur [16,17].

Let us assume that the refractive index of the medium of propagation is given by

$$n^2(\mathbf{r}) = n_0^2 + n_1^2 [\cos(\chi x) + \sin(\chi y + \beta)] \; , \tag{29}$$

where $n_1^2 \ll n_0^2$ and $\beta \in (0, 2\pi)$ is a constant parameter. From now on we assume that parameters of the system are away from the resonance ones, and geometrical optics and perturbation methods are applicable. The approximations in use are certainly applicable when the sample is not too long, and the backscattering is negligible. The condition of validity of the geometrical optics approximation, Eq. (14), means in the case Eq. (29) that the period of the medium inhomogeneity should be much larger than the wavelength of the radiation:

$$\frac{\chi}{n_0 k_0} \ll 1 \; . \tag{30}$$

To calculate ray deflections (26) we first find the "zero-approximation" ray trajectory, which is determined by geometrical optics equations (25) with the last term in the second equation omitted. Substitution of Eq. (29) in (25) yields

$$\dot{\mathbf{p}} = \frac{n_1^2 \chi}{2}[-\sin(\chi x), \cos(\chi y + \beta), 0] \; , \quad \dot{\mathbf{r}} = \mathbf{p} \; . \tag{31}$$

We solve this system of non-linear differential equations by the perturbation method under assumption that the variations of the refractive index are small in amplitude:

$$\delta = \frac{n_1^2}{n_0^2} \ll 1 \; . \tag{32}$$

If in zero order approximation in $\delta$ the ray has the momentum (wave vector) $\mathbf{p}_0 = (p_{x0}, p_{y0}, p_{z0})$ and crosses the origin $\mathbf{r} = 0$, then in the first order approximation in $\delta$ the solution of Eq. (31) gives the following trajectory of the wave in $\mathbf{p}$-space:

$$p_x = p_{x0} + \frac{n_1^2}{2p_{x0}}\cos(\chi p_{x0} s), \quad p_y = p_{y0} + \frac{n_1^2}{2p_{y0}}\sin(\chi p_{y0} s + \beta), \quad p_z = p_{z0}. \tag{33}$$

Deflections of circularly polarized rays can be found directly from Eq. (26) by substituting (33) and subsequent integration. It also can be estimated using rather convenient geometrical method proposed in [17], where it has was shown that, in the case of closed in **p**-space trajectories, the deviation of the circularly polarized ray is equal to the **p**-gradient of Berry phase, i.e.

$$\delta \mathbf{r}^\pm = -k_0^{-1}\frac{\partial \theta_B}{\partial \mathbf{p}}. \tag{34}$$

Berry phase grows with the ray length, and its increment during one period of the cyclic evolution in **p**-space is equal to the flux of the field $\mathbf{F}^\pm$ through contour $C$, Eqs. (21), (22), (28), i.e. to the solid angle $\Omega$ (with the corresponding sign), at which contour $C$ is seen from the origin of **p**-space (see Eq. 28 and [1,4]).

Let us analyze the rays located (in zero geometrical optics approximation) in $(x, y)$ plane. To do this we assume that $p_{z0} \to 0$ is infinitely small value (it is necessary for differentiation in Eq. (34)), and introduce the angle of propagation, $\alpha$, as the angle between $\mathbf{p}_0$ and $x$ axis:

$$\tan\alpha = \frac{p_{y0}}{p_{x0}}. \tag{35}$$

If the angle $\alpha = \pi/4$, i.e. $p_{x0} = p_{y0} = p_0/\sqrt{2}$, according to Eq. (33) the trajectory $C$ in momentum space is a circle in $(p_x, p_y)$ plane with the centre at the point $\mathbf{p}_0$ and radius $n_1^2/\sqrt{2}p_0$ (Fig. 1a) (So far we have assumed that $\beta = 0$ in Eq. (33)). At infinitely small $p_{z0}$ this circle is seen from the origin of **p**-space as an ellipse with the semi-axes $\rho_1 = n_1^2/\sqrt{2}n_0$ and $\rho_2 = n_1^2 p_{z0}/\sqrt{2}n_0^2$. The area of this ellipse equals $\pi\rho_1\rho_2 = n_1^4 p_{z0}/2n_0^3$, while the solid angle equals $\Omega = \pi n_1^4 p_{z0}/2n_0^5$. Hence, the Berry phase per one period of the inhomogeneity is equal to

$$\theta_{B0} = \mp \frac{\pi n_1^4 p_{z0}}{2n_0^5}. \tag{36}$$

In one period the ray runs the length $l_0 = 2\sqrt{2}\pi/\chi$, and after a number of a periods (when the deflections become noticeable) Berry phase becomes approximately

$$\theta_B \approx \theta_{B0}\frac{l}{l_0} = \mp\frac{n_1^4 \chi p_{z0}}{4\sqrt{2}n_0^5}l, \tag{37}$$

where $l$ is the total length of the ray. By substituting Eq. (37) into Eq. (34) we obtain

$$\delta z^\pm \approx \pm\frac{k_0^{-1}\chi n_1^4}{4\sqrt{2}n_0^5}l. \tag{38}$$

Note, that in spite of the fact that Berry phase, Eqs. (36), (37), equals zero at $p_{z0} = 0$, their momentum gradient, Eqs. (34), (38), is a finite quantity (compare with examples in [17]).

Eq. (38) means that right-hand and left-hand circularly polarized rays propagating in $(x, y)$ plane at $\alpha = \pi/4$ are deflected in the opposite $z$-directions, i.e. the polarization topological transport of electromagnetic waves in periodical medium, Eq. (29), takes place. If electromagnetic wave has mixed polarization, the ray will split into two circularly polarized eigen rays [16–18]. The effect, being of order of the small parameter $k_0^{-1}\chi$, is nevertheless proportional to length $l$ of the trajectory, and therefore can be significant at large distances.

It should be noted that two-dimensionally inhomogeneous medium under consideration possesses TE and TM exact eigen modes. However linear birefringence of the TE and TM modes of the system is proportional to the second power of the small geometrical optics parameter (14) (see, for instance, [27], and thus should be neglected in the approximation in use. We emphasize also that the reflection symmetry of the medium and of the TE and TM eigen modes with respect to the $z = 0$ plane does not contradict to the deflections of the rays in $z$ directions. The matter is that there is no transverse transport of the eigen modes. However their interference in the course of ray propagation leads to the effective deflection of the circularly polarized ray (which is superposition of a number of modes). This is connected with the reflection asymmetry for the circularly polarized rays. Indeed, reflection transformation changes the sign of the helicity. Maxwell equations are also not symmetrical with respect to the right and left circular polarizations (mind opposite signs of gauge fields for right and left polarizations in Eqs. (19), (21)). Thus these asymmetries cause topological spin transport under consideration.

Propagation of light in 2D photonic crystals has been considered in recent paper [19]. It was shown that a smooth inhomogeneity superimposed on the periodic crystal structure could induce topological spin transport of *eigen modes*. In distinction from [19] we have shown that polarization transport of *rays* can exist in perfectly periodic system itself.

To investigate in details the dependence of the ray deflection on the angle of propagation, $\alpha$, and on parameter $\beta$ (which characterises the position of the ray with respect to the periodic lattice), numerical calculation of the integral in Eq. (26) with Eq. (33) had been carried out. Figs. 2 show the deflection $\delta z^+$ of right-hand circularly polarized beam as a function of the angle of propagation, $\alpha$, at two samples of different lengths, $L$, with $\beta = 0$. The existence of the well-pronounced peak around $\alpha = \pi/4$ can be understood if one recalls that the deflection is proportional to the (oriented) area bounded by the one-period ray trajectory in $\mathbf{p}$-space. The trajectory for $\alpha = \pi/4$ is depicted in Fig. 1a. If $\tan\alpha = m$ (or $\cot\alpha = m$), where $m > 1$ is an integer number (in Figs. 1b and 1c $m = 2$ and $3$ respectively), the trajectories are closed lines, and the total oriented area of a one-period cycle equals zero, as well as Berry phase and ray deflection. At a longer sample, Fig. 2b, a number of resonant-like peaks at different angles $\alpha \neq \pi/4$ arise. These peaks occur at angles that correspond to non-commensurable values of $p_{x0}$ and $p_{y0}$, i.e. to irrational values of $\tan\alpha$. That is why the picture has random-like structure. Actually, the non-commensurability of $p_{x0}$ and $p_{y0}$ gives rise to breaking of time-reversal symmetry (this is necessary for Berry phase and accompanied phenomena) for ray under consideration. Indeed, the corresponding trajectories for such rays in $\mathbf{p}$-space are open, and in general enclose non-zero oriented areas. The area changes its sign at time-reversal transformation. In samples of finite length peaks occur at rational values of $\tan\alpha$ if $L$ is less than the period of wave trajectory in infinite medium. This condition gives the following estimate for the characteristic width of the peaks:

$$\delta\alpha \sim \frac{1}{\chi L} \ll 1 \ . \qquad (39)$$

In the limit $L \to \infty$ the peaks exist at irrational values of $\tan\alpha$. Amplitudes and widths of the peaks in Figs. 2 agree closely with Eq. (38) and estimate (39) respectively.

Figs. 3a–d present the dependences of the ray deflection $\delta z^+$ on the angle of propagation $\alpha$ at different values of parameter $\beta$. The maximum value of $\delta z^+$ decreases with increasing $\beta$ and changes sign at $\beta = \pi$. It should be noted that when $\beta = \pi/2$, the deflection $\delta z^\pm$ vanishes at $\alpha = \pi/4$, but occurs with the opposite signs just near the $\alpha = \pi/4$. It can be also understood by considering the wave trajectory in $\mathbf{p}$-space. At $\beta = \pi/2$ and $\alpha = \pi/4$ it is line (degenerated contour), and zero area corresponds to it. But when $\beta = \pi/2$ and $\alpha = \pi/4 + \gamma$ with $|\gamma| \ll 1$, the trajectory is an open curve (for finite $L$) that depicted qualitatively at Fig. 1d. Obviously it encloses non-zero oriented area, and the sign of this area correlates with the sign of $\gamma$ (compare with Fig. 3b).

Finally, let us discuss the possibility of the experimental observation of the predicted effect. If, for example, $\delta = 1/2$, Eq. (38) gives $\Delta z \equiv \delta z^+ - \delta z^- \approx 10^{-1} \lambda L / L_0$ where $\lambda = 2\pi / k_0 n_0$, and $L_0 = 2\pi / \chi$ is the period of the inhomogeneity. Hence, to get $\Delta z$ larger than the wavelength one needs a sample longer than ten inhomogeneity periods, $L > 10 L_0$. (Note that the lengthening of the sample leads, in accordance with Eq. (39), also to the narrowing of the peak, and can make it indiscernible). The splitting could be observed, for example, by using a converging (focused) beam whose width near the focal spot is of the order of a few wavelengths. When the splitting is too small to produce two separate beams of different polarization, the effect can be evidenced by measuring the polarization structure of the beam. Indeed, even small splitting of the 'mathematical' ray of the linear polarization into two circularly polarized rays causes the appearance of the opposite-sign circular polarizations at the opposite sides of the initially linearly polarized physical beam [28]. Another experimental possibility is to observe the propagation of a single circularly polarized beam, and to measure the deflection of its center of gravity, that can be done with the accuracy much higher than the width of the beam.

## 4. Conclusion

First observation of the polarization transport of electromagnetic waves in isotropic inhomogeneous medium is due to Zel'dovich et al. [14,15]. Different theoretical explanations for this "optical Magnus effect" have been suggested based on the mode approach, ray approximation, and spin-orbit interaction of photons. Recent theoretical investigations [16−20] show that this effect has pure topological nature and is closely related to Berry phase. In [16,17] it has been demonstrated in the framework of the classical geometrical optics, while in [18,19] the contemporary gauge field approach was used. Gauge field, Eqs. (21), (22), have been introduced in [18,19] out of a general reasoning, although without a full rigorous derivation. In [20] this field was obtained in a general form for relativistic quantum particles with arbitrary spin. Although the notion of gauge field for photons was mentioned a number of times in papers dealing with Berry phase theory, the first rigorous derivation of this field directly from Maxwell equations by general diagonalization and topological spin transport approach has been done in the present paper apparently for the first time. It can be said that the results of Sec. 2 present a complete adiabatic and semiclassical theory of the topological spin transport of photons and provide a rigorous grounds for the corresponding equations to be used.

To summarize, a semiclassical theory of the topological spin transport of photons in stationary inhomogeneous medium has been developed. The first step in the derivation is the diagonalization of the kinetic energy in Maxwell equations for monochromatic waves. A non-Abelian $\mathbf{U}(3)$ pure gauge potential in momentum space is induced, which in the adiabatic approximation breaks down into the product, $\mathbf{U}(3) \to \mathbf{U}(2) \times \mathbf{U}(1)$, where $\mathbf{U}(2)$ potential corresponds to transverse waves. Since photon is a massless particle, $\mathbf{U}(2)$ gauge potential is presented as a product of two Abelian potentials, $\mathbf{U}(2) \to \mathbf{U}(1) \times \mathbf{U}(1)$. These potentials describe the evolution of the right-hand and left-hand circularly polarized waves that are the photon eigen spin states with helicities $\pm 1$. The magnetic-monopole-like gauge field tensors corresponding to these potentials bring into existence a Lorentz-type force in the semiclassical equations of motion, and, hence, induce the topological spin transport of photons. As the result, the topological polarization ray deflection and splitting take place [16−18]. It was mentioned earlier [18,19], that it is analogous to the spin and anomalous Hall effects for electrons in solids.

As an application of the developed theory the topological polarization transport of electromagnetic waves in 3D medium with smooth periodic two-dimensional inhomogeneity is considered. By solving the geometrical optics ray equations perturbatively we show that a circularly polarized ray located in the plane of inhomogeneity experiences the topological polarization deflection orthogonal to this plane. The deflections are of the opposite signs for the left-hand and right-hand polarized rays, and are extremely sensitive to the angle of propagation. Examples of the dependences of the deflection on the angle of propagation and on the position of the ray with respect to the periodic structure are presented, and the physical implications of the revealed features are discussed.

Experimental observation of the effects caused by the topological polarization transport of light presents a real challenge, because they could mix up with the diffractional broadening of the physical ray (beam), which is known to be of the same order of smallness, i.e. proportional to the wavelength. When this broadening is larger than the topological ray splitting the topological polarization transport can induce right-hand and left-hand circular polarizations at opposite sides of the single physical ray [28], or can lead to the *depolarization* of the ray (compare with light depolarization that occurs due to the Berry phase increment [29,30]).

It should be also noted that the polarization deflection and splitting of rays are strongly connected with Berry phase inherent in circularly polarized waves. Therefore our results can be used in studying the Berry phase in periodic structure. For example, predicted increase of Berry phase (Eqs. (36), (37)) can be detected in a two-dimensional periodic structures (photonic crystals) by measuring the rotation of the wave polarization plane.

## Acknowledgment

The work was partially supported by INTAS (grant 03-55-1921), by Ukrainian President's Grant for Young Scientists GP/F8/51, and by Israeli Science Foundation (grant 328/02).

**Figure captions**

**Fig 1.** Wave trajectories in $(p_x, p_y)$ plane in the momentum space. Dimensionless units are used (see Eqs. (33)). The signs of oriented areas of the loops are shown.

**Fig 2.** Dependence of the deflection, $\delta z^+$, on the angle of propagation, $\alpha$, for two samples of different sizes. Fig. 2a corresponds to the sample of 40 periods of inhomogeneity, whereas Fig. 2b corresponds to the sample of 250 periods of inhomogeneity.

**Fig 3.** Same as in Fig. 2a for different $\beta$.

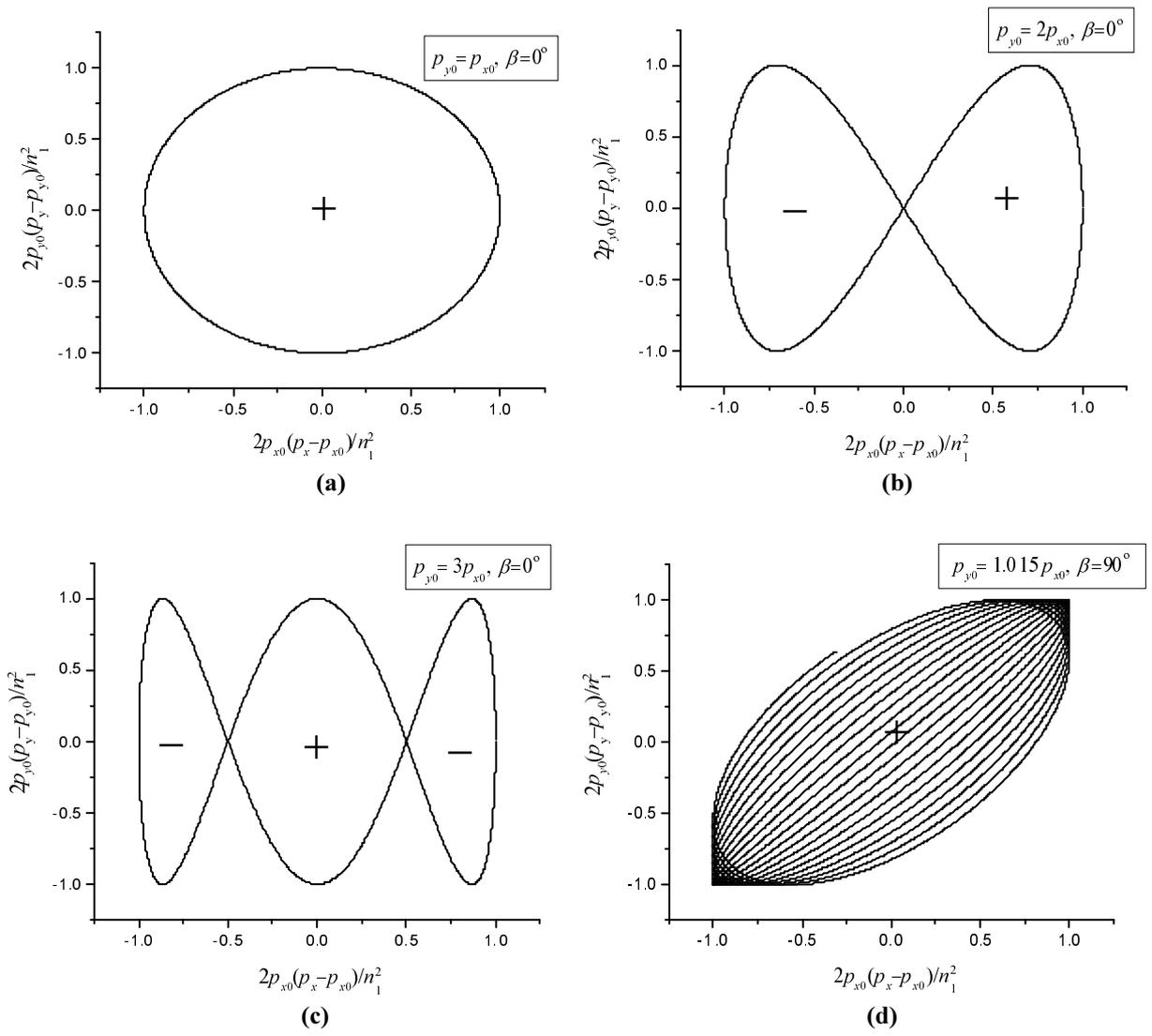

**Fig. 1**

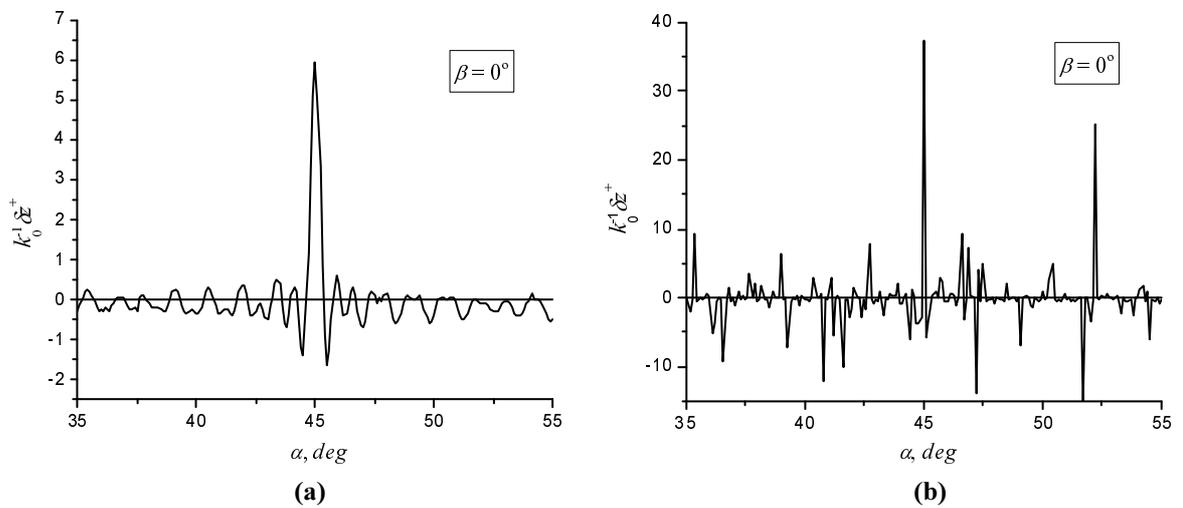

**Fig. 2**

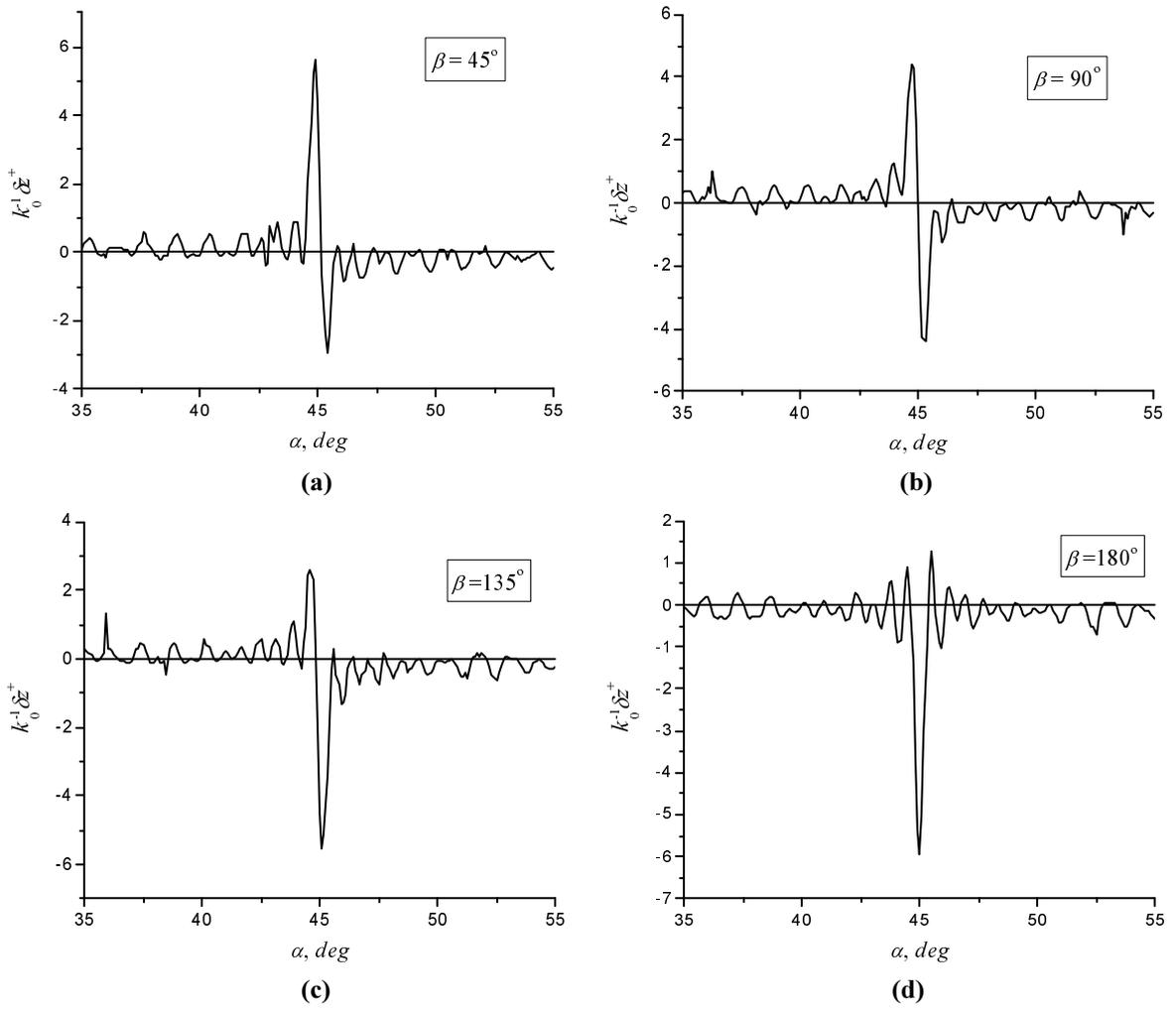

**Fig. 3**